\documentclass{article}
\usepackage{graphicx} 
\usepackage{hyperref}
\usepackage{amsmath,amsfonts,amssymb,amsthm,mathrsfs,tikz-cd, enumerate,mathtools,authblk,amscd,tikz-cd}

\DeclarePairedDelimiterX\set[1]\lbrace\rbrace{#1}

\newcommand{\ra}{\rightarrow}

\newcommand{\RR}{{\mathbb R}}

\newcommand{\ZZ}{{\mathbb Z}}

\newcommand{\QQ}{{\mathbb Q}}

\newcommand{\SAl}{{\mathscr A}_{\ell}}

\newcommand{\SU}{{\mathscr U}}

\newcommand{\mfkl}{{\mathfrak l}}

\newcommand{\mfkH}{{\mathfrak H}}
\newcommand{\mfkS}{{\mathfrak S}}

\newcommand{\bR}{{\mathfrak R}}
\newcommand{\bH}{\mfkH}
\newcommand{\bl}{\mfkl}
\newcommand{\bS}{\mfkS}

\newcommand{\HP}{{\mathsf H}}
\newcommand{\Ad}{{\rm Ad}}
\newcommand{\coker}{{\rm coker}}

\newcommand{\trho}{{\tilde\rho}}

\title{Higher symmetries and anomalies in quantum lattice systems}
\author{Anton Kapustin}
\author{Shixiong Xu}
\affil{\normalsize California Institute of Technology, Pasadena, CA 91125}

\begin{document}

\maketitle

\begin{abstract}
\noindent
    We define an 't Hooft anomaly index for a group acting on a 2d quantum lattice system by finite-depth circuits. It takes values in degree-4 cohomology of the group and is an obstruction to the on-site-ability of the group action. We introduce a 3-group (modeled as a crossed square) describing higher symmetries of a 2d lattice system and show that the 2d anomaly index is an obstruction for promoting a symmetry action to a morphism of 3-groups. This demonstrates that 't Hooft anomalies are a consequence of a mixing between ordinary symmetries and higher symmetries. Similarly, to any 1d lattice system we attach a 2-group (modeled as a crossed module) and interpret the Nayak-Else anomaly index as an obstruction for promoting a group action to a morphism of 2-groups. The meaning of indices of Symmetry Protected Topological states is also illuminated by higher group symmetry.
\end{abstract}
\pagebreak
\tableofcontents

\section{Introduction}

Quantum Field Theory (QFT) is a notoriously subtle subject, with mathematical foundations that remain unsettled. One could argue that the heart of QFT and the source of many of its difficulties lies in the concept of locality. The Haag-Kastler algebraic approach \cite{HaagKastler} represents the most systematic effort to build locality into QFT from the ground up. However, it is becoming increasingly clear that this framework falls short in certain respects.

For instance, it is now widely accepted that any local QFT must account for higher or generalized symmetries. Yet, existing mathematical formulations of QFT do not seem to adequately capture this notion. At a basic level, higher symmetries are those that act trivially on all local observables but can have nontrivial effects on extended ones \cite{gensym}. In a local QFT in $d$ spatial dimensions, symmetries should therefore not be described by a single group, but rather by a collection of groups $G_0,\ldots,G_d$, where $G_k$ encodes the symmetries acting on observables supported on $k$-dimensional submanifolds.

But this is only part of the story. The full symmetry structure of a local QFT is believed to be governed not by a list of groups, but by a $(d+1)$-group, a concept originating in homotopy theory. Why the combination of symmetry and locality leads to homotopy-theoretic structures remains a mystery.

Another area where our limited understanding of locality is keenly felt is the study of anomalies. Physicists often describe 't Hooft anomalies as obstructions to promoting a global symmetry of a QFT to a local one. However, the precise mathematical meaning of this statement is still unclear. It is natural to suspect that higher symmetries and anomalies are closely related, as both arise from the intricate interplay between symmetry and locality.

In this paper, we investigate the problem in the mathematically controlled setting of quantum lattice systems in one and two spatial dimensions. The structure of ’t Hooft anomalies in one-dimensional lattice systems is well understood. For symmetry actions implemented via finite-depth circuits, Nayak and Else \cite{NayakElse} showed that such anomalies are classified by a degree-3 cohomology class of the symmetry group 
$G$. It is also known that these anomalies have physical consequences akin to those encountered in quantum field theory \cite{KapSop}. The vanishing of the Nayak–Else anomaly is both a necessary and sufficient condition for the symmetry action to be “on-siteable,” a notion closely related to locality \cite{WS,Bolsetal}.

However, connecting lattice 't Hooft anomalies to higher symmetries presents a challenge: in the absence of Gauss law constraints, lattice systems lack genuinely nonlocal observables, and thus appear to lack microscopic higher symmetries.

Our starting point is the observation that any quantum lattice system in 
$d$ spatial dimensions possesses a natural 
$d$-form symmetry, namely $G_d=U(1)$. The charge associated with this symmetry is simply the phase of a local scalar observable. We demonstrate that in one- and two-dimensional systems, this apparently trivial higher-form symmetry combines with ordinary (0-form) symmetries (automorphisms of the algebra of local observables) into a nontrivial $(d+1)$-group. This mixing of ordinary and 
$d$-form symmetries leads to anomalies. In one dimension, this anomaly coincides with the Nayak–Else index. In two dimensions, we show that, assuming the vanishing of a certain other anomaly, it is captured by a cohomology class in 
$H^4(G,U(1))$.

In more detail, we propose that fully local symmetries of a quantum system in 
$d$ spatial dimensions are described by a connected homotopy 
$(d+1)$-type, i.e., a space $X$, defined up to homotopy, whose homotopy groups 
$\pi_n(X)$ are nontrivial only for $1\leq n\leq d+1$.
The group $\pi_n(X)$ is interpreted as the 
$(n-1)$-form symmetry group $G_{n-1}$. A naive symmetry action by an abstract group $G$ (one that does not account for locality) is a homomorphism 
$G\ra G_0=\pi_1(X)$. In contrast, a fully local action corresponds to a continuous map (up to homotopy) $BG\ra X$, where 
$BG$ is the classifying space of $G$. Such a map induces a homomorphism 
$\pi_1(BG)=G\ra \pi_1(X)=G_0,$ but not every group homomorphism arises from a map $BG\ra X$. A ’t Hooft anomaly is precisely the obstruction to constructing such a map that realizes a given homomorphism $G\ra G_0$. This formalizes the idea that higher symmetries encode locality and that anomalies obstruct fully local implementations of symmetry.

This framework also provides insight into invariants of symmetry-protected topological (SPT) states. A short-range entangled state that is invariant under 
$G$ naturally defines a way to localize the 
$G$-action. SPT invariants distinguish inequivalent such localizations.

Where does the homotopy type $X$ come from? Another key point, well known to mathematicians but less familiar to physicists, is that constructing a homotopy type does not require a topological space. Homotopy types can also be modeled algebraically or combinatorially. From this perspective, a connected homotopy 
$(d+1)$-type is equivalent to a 
$(d+1)$-group. A 1-group is simply an ordinary group, while a 2-group is a group-like monoidal category: a monoidal category in which all objects and morphisms are invertible. Given a $(d+1)$-group, one can construct its classifying space, 
$(d+1)$-type, though this step is often unnecessary. In this paper, we identify 2-groups and 3-groups that describe the symmetries of one- and two-dimensional lattice systems, and we show that ’t Hooft anomalies correspond to the fact that these higher symmetry groups are not decomposable into direct products of 0-form and higher-form symmetries—technically, they possess nontrivial Postnikov classes.

The structure of the paper is as follows. In Section 2, we review the definitions and basic properties of Quantum Cellular Automata and quantum circuits, with an emphasis on their localization features. Section 3 introduces the commutator pairing for circuits localized in quasi-one-dimensional regions. In Section 4, we use this pairing to construct an anomaly index for a group 
$G$ acting via circuits on a 2d lattice system, an analogue of the Nayak–Else index in two dimensions. The resulting expressions are quite intricate and would be difficult to guess without the higher symmetry perspective. Sections 5 and 6 provide this interpretation: first for the Nayak–Else index, and then for the 2d anomaly. The relevant algebraic structures, 2-groups and 3-groups, appear in the form of crossed modules \cite{Whitehead} and crossed squares \cite{Conduche1}, respectively. We also discuss an explicit example of a group action with a nontrivial anomaly in degree-4 cohomology. Finally, in Section 7, we outline how invariants of SPT phases can be understood through the lens of higher group symmetries.

We would like to thank Nikita Sopenko and Thomas Dumitrescu for discussions. We are especially grateful to Yu-An Chen for emphasizing a possible connection with the results of \cite{PRX} and for pointing out an example of a group action with a nontrivial anomaly. This example was included in the second version of the paper. This work was supported in part by the U.S.\ Department of Energy, Office of Science, Office of High Energy Physics, under Award Number DE-SC0011632 and by the Simons Investigator Award.

While this paper was in the final stages of preparation, we learned that similar results were obtained by Czajka, Geiko, and Thorngren \cite{Ryanetal} and Kawagoe and Shirley  \cite{WilburKyle}.

\section{QCAs and circuits}

Let $\alpha$ be an automorphism of the algebra of local observables $\SAl$. We say that $\alpha$ has range $r\in [0,+\infty)$ if for any $a\in\SAl$ supported on a site $j$,  $\alpha(a)$ is supported on a ball of radius $r$ with center at $j$. In this paper we will only consider automorphisms for which such an $r$ exists; they are called QCAs (Quantum Cellular Automata). The number $r$ is called the range of the QCA. QCAs form a subgroup of the group of automorphisms of $\SAl$.

A partitioned unitary of range $r$ is a collection of a finite or infinite number of unitary local observables with disjoint supports such that the diameter of the support of each unitary does not exceed $r$. The number $r$ is called the range of the partitioned unitary. Conjugation with a partitioned unitary of range $r$ is a QCA of range $r$.

A (unitary) circuit of depth $N$ is an $N$-tuple of partitioned unitaries. We will refer to these partitioned unitaries as the layers of a circuit. The range of a circuit is the sum of the ranges of its layers. Every circuit gives rise to a QCA: the product of QCAs corresponding to its layers. We will denote the QCA corresponding to a circuit $B$ by $\phi(B)$. 

It is important to distinguish circuits from the QCAs they generate, since a given QCA can be generated by many different circuits. Contrariwise, some QCAs cannot be generated by any circuit. In the case of 1d QCAs there is a simple necessary and sufficient condition which ensures that a QCA can be generated by a circuit: the vanishing of the GNVW index \cite{GNVW}. The GNVW index of a 1d QCA $\alpha$ is the value of a GNVW homomorphism $ind$ from the group of 1d QCAs to the group of positive rational numbers $\QQ_+$, with the group operation being multiplication. For example, the GNVW index of a translation of a chain of quidits by one site is $d^{\pm 1}$, hence the translation automorphism is not generated by a circuit. 

A concatenation of two circuits $B,B'$ of depth $N$ and $M$ gives a circuit $B\cdot B'$ of depth $N+M$. The concatenation operation is associative and makes the set of circuits into a semigroup. The map from circuits to QCAs is a semigroup homomorphism, i.e. $\phi(B\cdot B')=\phi(B) \phi(B')$ for any two circuits $B,B'$.

In what follows an important role is played by the notion of approximate localization. We say that a QCA $\alpha$ is approximately localized on a set $\Gamma\subset\RR^d$  if $\alpha$ acts trivially on all local observables supported farther from $\Gamma$ than some $r>0$. This notion of localization is "fuzzy": being approximately localized on $\Gamma$ is the same thing as being localized on any thickening of $\Gamma$, where by a thickening we mean all points of $\RR^d$ which are within some distance $s$ from $\Gamma$. In this paper we only deal with approximate localization, so in what follows we will omit the adjective "approximate". Similarly, a circuit is localized on $\Gamma\subset\RR^d$ if the supports of all unitaries which constitute its layers are within a distance $r<\infty$ from $\Gamma$. We will denote the group QCAs localized on $\Gamma$ and the semigroup of circuits localized on $\Gamma$ by $QCA_\Gamma$ and $Cir_\Gamma$, respectively. The group of QCAs generated by circuits localized on $\Gamma$ (i.e. the image of $Cir_\Gamma$ under the semigroup homomorphism $\phi$) will be denoted by $QCA^c_\Gamma$. $QCA^c_\Gamma$ is a normal subgroup in $QCA_\Gamma$. In particular, if $\Gamma$ is a straight line in $\RR^d$, $QCA^c_\Gamma$ is the kernel of the GNVW homomorphism $ind:QCA_\Gamma\ra\QQ_+$.

One can formalize the notion of approximate localization by introducing a kind of order (technically, a {\it pre-order}) on the set of subsets of the lattice \cite{AKY}. We say that $\Gamma\leq\Gamma'$ if $\Gamma$ is contained in some thickening of $\Gamma'$. If this is the case, then a QCA or a circuit localized on $\Gamma$ is also localized on $\Gamma'$. It may happen that $\Gamma\leq\Gamma'$ and $\Gamma'\leq\Gamma$, but $\Gamma\neq\Gamma'$ (this is why the relation $\leq$ is a called a pre-order rather than a partial order). For example, $\ZZ\leq\RR$ and $\RR\leq\ZZ$. As far as localization properties are concerned, such $\Gamma$ and $\Gamma'$ are equivalent. Note that if $\Gamma\leq \Gamma'$, then $QCA_\Gamma$ is a normal subgroup of $QCA_{\Gamma'}$, so $QCA_{\Gamma'}/QCA_{\Gamma}$ is a group. 

If a circuit is localized on $\Gamma$, then the corresponding QCA is also localized on $\Gamma$. The converse need not be true. In fact there are examples of QCAs which are generated by circuits and are localized on a proper subset of $\ZZ^d$ but cannot be generated by any circuit localized on this subset. The simplest examples arise from chiral Floquet phases in 2d \cite{CF1,CF2,CF3}. Such a phase is constructed using a 2d circuit which has the following property: its truncation to a half-plane $\HP$ generates a QCA which is localized on the boundary of $\HP$ and whose restriction to some thickening of $\partial\HP$ is a translation. Since translations have a nonzero GNVW index, such a QCA cannot arise from a circuit localized on $\partial\HP$.

In 1d the situation is simpler. Up to thickening, there are only three inequivalent proper subsets of $\RR$ to consider: left half-line, right half-line, and the origin. A QCA localized on the left/right has a vanishing GNVW index and thus can be generated by a circuit localized on the left/right. Consequently, a QCA localized at the origin can be generated by a circuit localized at the origin, i.e. it is a conjugation by a local unitary. This unitary is defined up to multiplication by a scalar of absolute value $1$. 

In this paper we only consider circuits localized on closed conical regions in $\RR^d$. If $\Gamma$ is a cone and $A$ is a circuit, there exist circuits $A_\Gamma$ and $B_\Gamma$ localized on $\Gamma$ and $\RR^d\backslash \Gamma$, respectively, such that $A=A_\Gamma\cdot B_\Gamma$. We will refer to $A_\Gamma$ as a truncation of $A$ to $\Gamma$. Truncations are not uniquely defined, but it is easy to see that any two truncations to $\Gamma$ differ by a circuit which is localized on the boundary of $\Gamma$. Similarly, if $\alpha$ is a QCA, we say that $\alpha_\Gamma\in QCA_\Gamma$ is a truncation of $\alpha$ to a closed cone $\Gamma$ if there exists a QCA $\beta_\Gamma$ localized on  $\RR^d\backslash\Gamma$ such that $\alpha=\alpha_\Gamma\beta_\Gamma$. Any two truncations of a QCA to a closed cone $\Gamma$ differ by a QCA localized on the boundary of $\Gamma$. For a general QCA, a truncation need not exist. But if $\alpha$ is generated by a circuit $A$, a truncation to $\Gamma$ always exists, as one can let $\alpha_\Gamma=\phi(A_\Gamma)$.

\section{The commutator pairing}

Let $\alpha,\beta$ be 1d QCAs localized on the left and right, respectively. Then $[\alpha,\beta]=\alpha\beta\alpha^{-1}\beta^{-1}$ is a QCA localized at the origin. Thus, there exists a unitary $u\in\SAl$ such that $[\alpha,
\beta]=\Ad_u$. Here by $\Ad_u$ we denote the automorphism $a\mapsto u a u^{-1}$. A priori $u$ is defined up to multiplication by a scalar.

For any $\alpha,\beta$ as above we are going to define a distinguished $u\in\SU_l$ in its equivalence class such that $[\alpha,
\beta]=\Ad_u$. This will give us a map $h:QCA_L\times QCA_R\ra \SU_l$ whose value  we will denote $\eta(\alpha,\beta)$. 

Suppose $\alpha=\phi(A)$ and $\beta=\phi(B)$, where $A$ is a left-localized circuit and $B$ is a right-localized circuit. We will begin by defining two functions $\eta_R:Cir_L\times Cir_R\ra \SU_l$ and $\eta_L:Cir_L\times Cir_R\ra \SU_l$. 
The first of these, $\eta_R(A,B)$, is defined by recursion on the number of layers of $B$. For a single layer $B$, we let
$$
\eta_R(A,B)=\lim_{r\ra\infty}\alpha(B_r) B_r^{-1},
$$
where $B_r\in\SU_l$ is obtained from $B$ by replacing with $1$ all unitary observables whose support is farther than $r$ from the origin. 
It is easy to see that the limit exists. In fact, the unitary observable $\alpha(B_r)B_r^{-1}$ is constant for $r$ larger than the range of $\alpha$. Note also that with this definition for any right-localized single-layer $B$ and any $a\in\SAl$ we have 
\begin{align}\label{eq:Adproperty}
\Ad_{\eta_R(A,B)}(a)=[\alpha,\beta](a).
\end{align}
Indeed, assuming a sufficiently large $r$ we have
\begin{multline}
\eta_R(A,B) a \eta_R(A,B)^{-1}=\alpha(B_r) B_r^{-1} a B_r\alpha(B_r)^{-1}=\alpha(B_r)\beta^{-1}(a)\alpha(B_r)^{-1}\\
=\alpha(B_r \alpha^{-1}\beta^{-1}(a)B_r^{-1})=[\alpha,\beta](a).
\end{multline}

If $\eta_R(A,B)$ has been defined for all right-localized circuits $B$ with $N$ layers, then $\eta_R(A,B\cdot B')$, where $B'$ is a right-localized single-layer circuit, is defined to be
\begin{align}\label{eq:rightmult}
    \eta_R(A,B\cdot B')=\eta_R(A,B)\ {}^{\phi(B)} \eta_R(A,B'),
\end{align}
where for any $u\in\SU_l$ and any QCA $\gamma$ we denote ${}^\gamma u:= \gamma(u)$ (this notation is introduced to avoid a proliferation of parentheses). It is easy to prove by induction on the number of layers that with this recursive definition the identity (\ref{eq:Adproperty}) holds for an arbitrary right-localized circuit $B$ while the identity (\ref{eq:rightmult}) holds for arbitrary right-localized circuits $B,B'$. 

We also note the following property of the function $h$:
\begin{align}\label{eq:leftmult}
\eta_R(A\cdot A',B)={}^{\phi(A)} \eta_R(A',B)\ \eta_R(A,B).
\end{align}
It can also be proved by induction on the number of layers of $B$.

The function $\eta_L:Cir_L\times Cir_R\ra\SU_l$ is defined by recursion on the number of layers of $A\in Cir_L$. For a single-layer $A$ we let
$$
\eta_L(A,B)=\lim_{r\ra\infty} A_r \beta(A_r^{-1}). 
$$
Having defined $\eta_L(A,B)$ for all $N$-layered $A$, we define it for an $(N+1)$-layered circuit $A\cdot A'$, where $A'\in Cir_L$ is single-layered by demanding 
\begin{align}
    \eta_L(A\cdot A',B)={}^{\phi(A)} \eta_L(A',B)\ \eta_L(A,B).
\end{align}
Now one can check that this identity remains true if $A$ and $A'$ are arbitrary left-localized circuits. The function $\eta_L$ also satisfies the other two identities satisfied by $\eta_R$ (namely, (\ref{eq:Adproperty}) and (\ref{eq:rightmult}). This implies that $\eta_R(A,B)=\lambda_{AB} \eta_L(A,B)$ for some $\lambda_{AB}\in U(1).$ In fact, $\eta_L(A,B)=\eta_R(A,B)$ for all $A,B$. Indeed, when both $A$ and $B$ are single-layered, for sufficiently large $r$ and $s$ we have
$$
\eta_L(A,B)=A_r \beta(A_r^{-1})=A_r B_s A_r^{-1} B_s^{-1}=\alpha(B_s) B_s^{-1}=\eta_R(A,B).
$$
Since they satisfy the same identities (\ref{eq:rightmult}) and (\ref{eq:leftmult}), recursion on the number of layers of $A$ and $B$ shows that $\eta_L=\eta_R$. 

Finally, it is clear from the definitions that $\eta_R$ depends only on $\alpha=\phi(A)$ and $B$, while $\eta_L$ depends only on $A$ and $\beta=\phi(B)$. Therefore they both depend only on $\alpha$ and $\beta$, not on the circuits used to generate them. We are entitled to denote their common value $\eta(\alpha,\beta)\in\SU_l$. The function $\eta:QCA_L\times QCA_R\ra \SU_l$ satisfies 
\begin{align}
    \Ad_{\eta(\alpha,\beta)}&=[\alpha,\beta],\\
    \eta(\alpha,\beta\beta')&=\eta(\alpha,\beta)\ {}^{\beta} \eta(\alpha,\beta'),\\
    \eta(\alpha\alpha',\beta)&={}^{\alpha} \eta(\alpha',\beta)\  \eta(\alpha,\beta).
\end{align}

We also note the following identities which follow directly from the definitions of $\eta_L$ and $\eta_R$:
$$
\eta(\Ad_u,\beta)=u\beta(u^{-1}),\quad \eta(\alpha,\Ad_u)=\alpha(u)u^{-1}.
$$
Here $u\in\SU_l$, $\alpha\in QCA_L$ and $\beta\in QCA_R$ are  arbitrary.

Finally, let $\gamma$ be an arbitrary QCA. Then 
$$
\eta(\gamma\alpha\gamma^{-1},\gamma\beta\gamma^{-1})={}^\gamma \eta(\alpha,\beta).
$$
This also follows easily from the definitions of $\eta_L$ and $\eta_R$.

\section{An anomaly index in two dimensions}

Suppose we are given a homomorphism from an abstract group $G$ to $QCA^c_{\RR^2}$. In this section we will define, under certain assumptions, an anomaly index for this action which takes values in $H^4(G,U(1))$. The definition will depend on a choice of a point of $S^1$ "at infinity", or equivalently, on a choice of a half-plane $\HP\subset\RR^2$; we believe that the index is independent of this choice but will not try to prove it.

By a choice of coordinates, we may identify $\HP$ with the lower half-plane in $\RR^2$ and identify its boundary $\partial\HP$ with $\RR$. We denote the left and right half-lines of the latter $L$ and $R$, respectively. For any quasi-1d $X\leq\ZZ^2$ (i.e for any $X$ which is equivalent to a line or a half-line in $\RR^2$) we will denote by $QCA^c_X$ the subgroup of $QCA_X$ consisting of elements with a vanishing GNVW index. By the properties of the GNVW index, $QCA^c_X$ is a normal subgroup of $QCA_X$. Moreover, if $X\leq Y$, then $QCA^c_X$ is a normal subgroup of $QCA_Y$. When $X$ if a half-line, the theory of the GNVW index tells us that $QCA^c_X=QCA_X$. On the other hand, we know that $QCA_{\partial\HP}$ is strictly larger than $QCA^c_{\partial\HP}$. 

Suppose we are given an action of a group $G$ by circuits, i.e. a homomorphism $\rho_0:G\ra QCA^c_{\RR^2}$. Truncating these circuits to $\HP$ we get a map $\tilde\rho:G\ra QCA^c_{\HP}$. It is defined up to $\trho(g)\mapsto \gamma(g)\trho(g)$, where $\gamma:G\ra QCA_{\partial\HP}$ is an arbitrary function. $\trho$ is not a homomorphism, but satisfies
$$
\mu(g,h)=\tilde\rho(g)\tilde\rho(h)\tilde\rho(gh)^{-1}\in QCA_{\partial\HP}. 
$$
We will say that $\rho_0$ is on-siteable if for any $\HP$ one can choose $\trho$ so that it is a homomorphism, i.e. so that $\mu(g,h)=1$ for all $g,h$.
The first obstruction to on-siteability of $\rho_0$ is a class in $H^2(G,\QQ_+)$ \cite{Zhangetal,Elseetal}. Indeed, using the properties of the GNVW index it is easy to see that  $ind(\mu(g,h))$ is a 2-cocycle on $G$ with values in $\QQ_+$ whose cohomology class is independent of the choice of the map $\trho$. If this 2-cocycle is not a coboundary, no modification of $\trho$ by QCAs localized on $\partial\HP$ can make all $\mu(g,h)$ belong to the subgroup $QCA^c_{\partial\HP}$, much less make it trivial. Contrariwise, if $ind(\mu(g,h))$ is a coboundary, then such a redefinition of $\tilde\rho$ can be accomplished. 

Suppose the cohomology class of $ind(\mu)$ vanishes.  Then it is possible to define a further class valued in $H^4(G,U(1))$ which obstructs the on-siteability of $\rho_0$. First, we choose a truncation $\tilde\rho:G\ra QCA_{\HP}$ such that $\mu(g,h)\in QCA^c_{\partial\HP}$ for all $g,h\in G$. Then we can find $\alpha(g,h)\in QCA_L$ and $\beta(g,h)\in QCA_R$ such that $\mu(g,h)=\alpha(g,h)\beta(g,h)$ for all $g,h\in G$. $\beta:G\times G\ra QCA_R$ is defined up to $\beta(g,h)\mapsto \Ad_{v(g,h)}\beta(g,h)$ where $v:G\times G\ra \SU_l$ is an arbitrary function. Of course, $\alpha$ is determined by $\beta$.

To make the formulas below more readable, for any two QCAs $\nu,\alpha$ we denote ${}^\nu\!\alpha=\nu\alpha\nu^{-1}$. Note that if $\alpha\in QCA_X$, then ${}^\nu\!\alpha\in QCA_X$, for all $\nu$.

Consider the expression
$$
\beta(g,h)\beta(gh,k)\beta(g,hk)^{-1}\, {}^{\tilde\rho(g)}\!\beta(h,k)^{-1}.
$$
It is easy to see that it is a QCA localized at the origin, thus there exists $u(g,h,k)\in \SU_l$ such that
$$
\Ad_{u(g,h,k)}=\beta(g,h)\beta(gh,k)\beta(g,hk)^{-1}\, {}^{\tilde\rho(g)}\!\beta(h,k)^{-1}.
$$
It is defined up to $u(g,h,k)\mapsto \psi(g,h,k) u(g,h,k)$ where $\psi:G\times G\times G\ra U(1)$ is an arbitrary function.

We are finally ready to write down a 4-cocycle $\tau$ on $G$ with values in $U(1)$. Consider the following function $\tau:G^4\ra \SU_l$: 
\begin{multline}\label{eq:tau}
\tau(g,h,k,l)=u(g,h,k)\,{}^{\trho(g)}\!\beta(h,k)\left(u(g,hk,l)\right)\trho(g)\left(u(h,k,l)\right)\\
{}^{\trho(g)\trho(h)}\!\beta(k,l)\left(u(g,h,kl)^{-1}\right)
\eta\left(\alpha(g,h),{}^{\beta(g,h)\trho(gh)}\!\beta(k,l)\right)
\beta(g,h)\left(u(gh,k,l)^{-1}\right)
\end{multline}
A somewhat lengthy but straightforward computation (see Appendix) shows that  $\Ad_{\tau(g,h,k,l)}=1$, therefore $\tau$ is a $U(1)$-valued 4-cochain on $G$. It is easy to see that the ambiguities in the choice of $u(g,h,k)$ change it only by a coboundary. A much lengthier computation, also sketched in the Appendix, shows that the ambiguities in the choice of $\trho(g)$ and $\beta(g,h)$ do not affect $\tau$. Finally, we will argue in Section 6 that $\tau$ is a 4-cocycle. Thus the cohomology class of $\tau$ is well-defined. Also, if there exists a truncation $\trho:G\ra QCA_{\HP}$ which is a homomorphism, then one can choose $\beta(g,h)=1$ for all $g,h\in G$ and $u(g,h,k)=1$ for all $g,h,k\in G$, hence $\tau(g,h,k,l)=1$ for all $g,h,k,l\in G$. Thus the cohomology class of $\tau$ is an obstruction for on-siteability of  $\rho_0$. 

\section{Higher symmetries and the Nayak-Else index}

\subsection{Crossed modules}

Our formula for the 2d anomaly index is rather obscure. In the next section we will interpret it in terms of a certain naturally occurring 3-group of symmetries. As a preparation, in this section we will interpret the 1d anomaly index of Nayak and Else in terms of a certain 2-group of symmetries. Our discussion here can be viewed as an elaboration of Appendix B of Ref. \cite{NayakElse}. 

The Nayak-Else index is defined for any group homomorphism $G\ra QCA^c_\RR$ and takes values in $H^3(G,U(1))$. Degree-3 group cohomology classes arise naturally from connected homotopy 2-types, or equivalently from 2-groups. A convenient model for a 2-group is a crossed module. Crossed modules were introduced by J.H.C. Whitehead  \cite{Whitehead}, see \cite{MacLanecategories} or \cite{BaezLauda} for a more pedagogical exposition. A crossed module is a pair of groups $M,N$, a homomorphism $\partial:M\ra N$, and an action of $N$ on $M$ by automorphisms. These data must satisfy two conditions:
\begin{align}
    \partial  ({}^n\!m)&=n(\partial m) n^{-1},\ \forall n\in N,\ \forall m\in M,\\
    {}^{\partial m_0}\!m_1&=m_0 m_1 m_0^{-1},\ \forall m_0,m_1\in M.
\end{align}
Here ${}^n\!m$ denotes the action of $n\in N$ on $m\in M$. The first condition ensures that the image of $\partial$ is a normal subgroup of $N$ and thus $\coker\partial=N/{\rm im}\partial$ is a group. The second condition ensures that $\ker\partial$ is an abelian group. 

The connected homotopy 2-type corresponding to a crossed module has $\pi_1=\coker\partial$ and $\pi_2=\ker\partial$. To determine the 2-type completely, one also needs to specify an element $\bl\in H^3(\pi_1,\pi_2)$ (the Postnikov invariant). When $\bl$ vanishes, the homotopy 2-type is the product of Eilenberg-MacLane homotopy types $K(\pi_1,1)\times K(\pi_2,2)$, otherwise it is a nontrivial fibration over $K(\pi_1,1)$ with fiber $K(\pi_2,2)$. The Postnikov invariant can be read off the crossed module as follows. First one chooses a map $\sigma:\pi_1\ra N$ such that $\sigma$ followed by the projection to $\pi_1$ is the identity map. Then one constructs $\nu:\pi_1\times\pi_1\ra N$ by letting
\begin{align}
\nu(\gamma,\gamma')=\sigma(\gamma)\sigma(\gamma') \sigma(\gamma\gamma')^{-1},\quad \gamma,\gamma'\in \pi_1. 
\end{align}
It is easy to see that $\nu$ followed by the projection $N\ra\pi_1$ is the trivial map (maps $\pi_1\times\pi_1$ to the identity element), hence there exists $\tilde\nu:\pi_1\times\pi_1\ra M$ such that $\partial\circ \tilde\nu=\nu$. $\tilde\nu$ is defined up to multiplication by a function on $\pi_1\times\pi_1$ with values in $\ker\partial$. Now form
\begin{align}\label{eq:ell}  \ell(\gamma,\gamma',\gamma'')=\tilde\nu(\gamma,\gamma')\tilde\nu(\gamma\gamma',\gamma'')\tilde\nu(\gamma,\gamma'\gamma'')^{-1}\, {}^{\sigma(\gamma)}\!\tilde\nu(\gamma',\gamma'')^{-1}.
\end{align}
It is easy to check that $\partial\circ \ell$ is the trivial map, hence $\ell$ is a map from $\pi_1\times\pi_1\times\pi_1$ to $\ker\partial$. One can check that it is a 3-cocycle and that its cohomology class $\bl$ does not depend on the choice of $\sigma:\pi_1\ra N$ and $\tilde\nu:\pi_1\times\pi_1\ra M$. 

In what follows, we will also make use of morphisms of crossed modules. A good definition of a morphism is somewhat subtle.  Morphisms between a pair of crossed modules naturally form a groupoid, i.e. a category where all morphisms have an inverse.\footnote{Actually, this groupoid  can be promoted to a 2-groupoid, but we will not need this richer structure.} The definition of the groupoid of morphisms is rigged so that isomorphism classes of morphisms between a pair of crossed modules are in bijection with homotopy classes of maps between the corresponding homotopy 2-types. An appropriate definition of a morphism between crossed modules is spelled out in \cite{Noohi}. We will only need the special case when the source crossed module has $M=1$ (i.e. it is simply a group $G$). A {\it weak morphism} from such a crossed module to a general crossed module $(M,N,\partial)$ is a group $E$ which is a nonabelian extension of $G$ by $M$, together with a group homomorphism $\phi:E\ra N$. These data must fit into a commutative diagram
$$
    \begin{tikzcd}
M \arrow[r] \arrow[dr, "\partial"'] & E \arrow[r] \arrow[d, "\phi"] & G \\
& N &
\end{tikzcd}
$$
Additionally, the embedding of $M$ into $E$ must be equivariant with respect to an action of $E$ on $M$. This action arises as a composition of the homomorphism $\phi$ and the action of $N$ on $M$.

Concretely, any such extension $E$ (which can be identified with $M\times G$ as a set) can be described by a function $\trho:G\ra N$ and a function $\mu:G\times G\ra M$ satisfying
\begin{align}\label{eq:1eq}
\trho(g)\trho(h)\trho(gh)^{-1}=\partial \mu(g,h),\quad g,h\in G
\end{align}
and 
\begin{align}\label{eq:2eq}
\mu(g,h)\mu(gh,k)\mu(g,hk)^{-1}{}^{\trho(g)}\!\mu(h,k)^{-1}=1.
\end{align}
Note that $\trho$ is not a homomorphism, but it does become a homomorphism $\rho:G\ra\coker\partial$ when composed with the projection $N\ra\coker\partial$. In terms of the data $(\trho,\mu)$, the group law on $E\simeq M\times G$ is given by
$$
(m_0,g_0)\cdot(m_1,g_1)=(m_0 {}^{\trho(g_0)}\! m_1 \mu(g_0,g_1),g_0 g_1).
$$
The homomorphism $\phi:E\ra N$ is given by
$$
\phi:(m,g)\mapsto \partial m\cdot\trho(g),\quad m\in M,g\in G.
$$

Note that for a fixed $\rho:G\ra\coker\partial$, the equation (\ref{eq:1eq}) can always be solved by some pair $(\trho,\mu)$, but the equation (\ref{eq:2eq}) is not automatic. The left-hand side of (\ref{eq:2eq}) is automatically a 3-cocycle on $G$ with values in $\ker\partial$, and it is easy to show that both equations can be solved iff this 3-cocycle is a coboundary. Comparing with (\ref{eq:ell}) we see that this 3-cocycle is the pull-back of the 3-cocycle $\ell$  via $\rho$. Hence a morphism from $G$ to the crossed module $(M,N,\partial)$ corresponding to a fixed $\rho:G\ra\coker\partial$ exists iff the pull-back of the Postnikov class via $\rho$ is trivial.

An isomorphism between two weak morphisms is an isomorphism of the corresponding groups $E$ and $E'$ which commutes with all the relevant maps to and from $E$ and $E'$ \cite{Noohi}. Isomorphisms can be described concretely as modifications of the data $(\trho,\mu)$ which preserve the equations (\ref{eq:1eq},\ref{eq:2eq}). If we fix the homomorphism $\rho:G\ra\coker\partial$ induced by $\trho$ and assume that the pull-back of the Postnikov class is trivial, then one can show that isomorphism classes of pairs $(\trho,\mu)$ are in bijection with elements of $H^2(G,\ker\partial)$ \cite{Noohi}. The bijection is not canonical but depends on a choice of a "basepoint" solution $(\trho_0,\mu_0)$. Concretely, suppose we are given a particular solution $(\trho_0,\mu_0)$ of the equations (\ref{eq:1eq},\ref{eq:2eq}). Then all other solutions inducing the same homomorphism $\rho:G\ra\coker\partial$ are isomorphic to solutions of the form  $(\trho_0,b\cdot \mu_0)$, where $b:G\times G\ra\ker\partial$ is a 2-cocycle, and two such solutions are isomorphic iff the corresponding 2-cocycles are cohomologous. 

From a homotopy-theoretic viewpoint, isomorphism classes of weak morphisms from $G$ to a crossed module $(M,N,\partial)$ are in bijection with homotopy classes of maps from $BG$ to the connected homotopy 2-type $X$ modeled by the crossed module \cite{Noohi}. Note that there is a map $X\ra B\pi_1(X)$ defined up to homotopy, so every map $BG\ra X$ also gives rise to a map $BG\ra B\pi_1(X)$. Fixing $\rho:G\ra \coker\partial=\pi_1(X) $ means that we fixed a group homomorpism $G\ra\pi_1(X)$, or equivalently a homotopy class of maps $BG\ra B\pi_1(X)$. If the pull-back of the Postnikov class of $X$ to $BG$ via $\rho$ is nontrivial, it is not possible to lift the chosen map $BG\ra B\pi_1(X)$ to a map $BG\ra X$. If the pull-back is trivial, then such maps do exist, but in general are not unique, even up to homotopy. It can be shown that homotopy classes of such maps are in bijection with homotopy classes of maps $G\ra K(\pi_2(X),2)$, i.e. with elements of $H^2(G,\ker\partial)$. This illustrates that the category of crossed modules and isomorphism classes of weak morphisms correctly models the category of connected homotopy 2-types.

\subsection{A crossed module for a 1d lattice system}

To every 1d quantum lattice system one can associate a crossed module  by letting $N=QCA_R$, $M=\SU_l$ and $\partial:u\mapsto \Ad_u$. The group $N$ acts on $M=\SU_l$ in the obvious way. The corresponding homotopy 2-type $\bR$ has $\pi_1=QCA_R/QCA_0$ and $\pi_2=U(1)$. Here $QCA_0$ is the group of QCAs localized at the origin. The corresponding Postnikov class $\bl_{\bR}$ is nonzero. Indeed, let $G$ be a group which acts on the 1d system by circuits. That is, suppose we are given a homomorphism $\rho_0:G\ra QCA^c_\RR$. Truncating it to the right half-line we obtain a homomorphism $\rho:G\ra QCA_R/QCA_0=\coker\partial$. By comparing the definition of the Nayak-Else class \cite{NayakElse} with the definition of the Postnikov class, one immediately sees that the Nayak-Else index of $\rho_0$ is precisely $\rho^*\bl$, i.e. the cohomology class of the 3-cocycle $\ell(\rho(g),\rho(h),\rho(k))$. Since there are  examples of group actions on 1d systems which have a nonzero Nayak-Else index, $\bl$ must be nonzero as well. Thus nontrivial higher groups appear already in the case of 1d lattice systems. 

The identification of the Nayak-Else index as the pull-back of the Postnikov class fits well with the philosophy explained in the introduction. An ordinary group $G$ can be regarded as a crossed module with $N=G$ and $M=\{1\}$. In other words, $G$ describes a connected homotopy 1-type, which can be viewed as a connected homotopy 2-type with $\pi_1=G$ and  $\pi_2=1$. On the other hand, kinematic symmetries of the system which are localized on the right half-line can be described by the crossed module $\bR$ defined in the previous paragraph. A local action of $G$ should give rise to a weak morphism of crossed modules $G\ra \bR$. Part of the data of such a morphism is a group homomorphism from $G$ to $\pi_1(\bR)=QCA_R/QCA_0$. Any truncation of the global $G$-action to the right half-line gives rise to such a homomorphism. But as explained in the previous subsection, it can be extended to a weak morphism $G\ra\bR$ if and only if the pull-back of the Postnikov class (i.e. the Nayak-Else anomaly) vanishes. Accepting  the interpretation proposed in the introduction, this means that the vanishing of the Nayak-Else anomaly is a necessary and sufficient condition for the localizability of the $G$-action to the half-line.


\section{Higher symmetries and the 2d anomaly index}

\subsection{Crossed squares}

To interpret the 2d anomaly index, we need the notion of a 3-group.\footnote{What we call a 3-group is usually called a weak 3-group. While every weak 2-group is equivalent to a strict 2-group, not every weak 3-group is equivalent to a strict one. In this paper we do not use strict 3-groups, so by a 3-group we will mean a weak 3-group.} 3-groups are essentially the same as connected homotopy 3-types, but described algebraically. There are several equivalent algebraic models of 3-groups. For our purposes, the most convenient one is the one given by crossed squares of groups \cite{crossedsquare}. A crossed square is a quadruple of groups and group homomorphisms which fit into a commuting diagram:
$$
\begin{CD}
L @>f>> M \\
@VgVV @VVvV \\
N @>u>> P
\end{CD}
$$
In addition $P$ acts on $L,M,N$ by automorphisms in such a way that the maps $f$ and $g$ are $P$-equivariant, and we have the following four crossed modules:
$$
\begin{CD}
M @>v>> P,\ N @>u>> P,\ L @>v\circ f>> P,\ \ L @>u\circ g>> P
\end{CD}
$$
Finally, part of the data of a crossed square is a map $\eta:M\times N\ra L$ which satisfies 
\begin{align}
f(\eta(m,n))&=m\, {}^{u(n)}\!m^{-1}, &g(\eta(m,n))&={}^{v(m)}\!n \, n^{-1},\\
\eta(f(l),n)&=l\,{}^{u(n)}\!l^{-1},  &\eta(m,g(l))&={}^{v(m)}\!l\, l^{-1},\\
\eta(m m',n)&={}^{v(m)}\!\eta(m',n)\eta(m,n), & \eta(m,nn')&=\eta(m,n)\, {}^{u(n)}\!\eta(m,n'),\\
\eta({}^p\!m,{}^p\!n)&={}^p\!\eta(m,n), & &
\end{align}
for all $m,m'\in M$, $n,n'\in N$ and $p\in P$.

Suppose we are given a crossed square. The homotopy groups of the corresponding homotopy 3-type $\mathfrak H$ can be extracted as follows. Note first that $N$ acts on $M$ via $n: m\mapsto {}^{u(n)}\! m $. Thus we can form a semidirect product $M\rtimes N$. Second, we define homomorphisms $\delta:L\ra M\rtimes N$ and $\partial:M\rtimes N\ra P$ by
$$
\delta:l\mapsto (f(l)^{-1},g(l)),\quad \partial:(m,n)\mapsto v(m)u(n).
$$
One can easily check that both $\delta$ and $\partial$ are group homomorphisms which satisfy $\partial\circ\delta=1$, thus we get a complex of nonabelian groups
$$
\begin{CD}
L @>\delta>> M\rtimes N @>\partial>> P
\end{CD}
$$
It can be also be checked that the images of both $\partial$ and $\delta$ are normal subgroups, so that the homology sets of the complex are groups. These homology groups are the homotopy groups of the 3-group  $\bH$:
$$
\pi_1(\bH)=\coker\partial,\quad \pi_2(\bH)=\ker\partial/{\rm im}\delta,\quad \pi_3(\bH)=\ker\delta.
$$

Of course, the homotopy groups do not determine the 3-type uniquely. When all three homotopy groups are nontrivial, the crossed square is an efficient way to encode all the data of the 3-type without introducing too much extraneous information. When only two homotopy groups are nontrivial, the situation simplifies. For example, when $\pi_2(\bH)$ is trivial, a complete description is provided by $\pi_1(\bH),\pi_3(\bH)$, and a Postnikov class ${\mathfrak t}\in H^4(\pi_1(\bH),\pi_3(\bH))$. Let us indicate how to extract this class from a crossed square.

As explained above, to every crossed square one can attach a normal complex of nonabelian groups. This complex has additional structure, that of a 2-crossed module, which gives yet another way to encode a connected homotopy 3-type \cite{Conduche2}. In general, a 2-crossed module consists of a normal complex of groups
$$
\begin{CD}
L @>\delta>> K @>\partial>> P,
\end{CD}
$$
an action of $P$ by automorphisms on both $L$ and $K$, and a map $\{\cdot,\cdot\}:K\times K\ra L$. The map $\{\cdot,\cdot\}$ is called a braiding or a Pfeiffer lifting and satisfies the following equations:
\begin{align}
    \delta\{k_0,k_1\}&=k_0 k_1 k_0^{-1}\, {}^{\partial k_0}\!k_1^{-1},\\
    \{\delta l_0,\delta l_1\}&=[l_0,l_1],\\
    \{\delta l,k\}\{k,\delta l\}&=l\, {}^{\partial k}l^{-1},\\
    \{k_0,k_1k_2\}&=\{k_0,k_1\}\{k_0,k_2\}\{\delta\{k_0,k_2\}^{-1},{}^{\partial k_0}\!k_1\},\\
    \{k_0k_1,k_2\}&=\{k_0,k_1k_2k_1^{-1}\}{}^{\partial k_0}\!\{k_1,k_2\},\\
    {}^p\{k_0,k_1\}&=\{{}^pk_0,{}^pk_1\}.
\end{align}
Here $l,l_0,l_1\in L$, $k,k_0,k_1,k_2\in K$ and $p\in P$ are arbitrary elements. 

Every crossed square gives rise to a 2-crossed module with $K=M\rtimes N$ \cite{Conduche2}. The Pfeiffer lifting is constructed\footnote{There appears to be an error in the formula for $\{\cdot,\cdot\}$ given in \cite{Conduche2}. This error was corrected in \cite{AU}. Note that the Pfeiffer lifting of \cite{AU} is the inverse of that in \cite{Conduche1,Conduche2}. We follow the convention of \cite{Conduche1,Conduche2}.} from $\eta:M\times N\ra L$:
$$
\{(m_0,n_0),(m_1,n_1)\}=\eta(m_0,n_0 n_1 n_0^{-1})^{-1}.
$$
It is straightforward but tedious to check that the map $\{\cdot,\cdot\}$ has all the required properties. 

It is instructive to look at the special case when $P$ is the trivial group and the map $\delta$ is also trivial. Then the first two equations force both $L$ and $K$ to be commutative, while the next three equations are equivalent to the statement that the Pfeiffer lifting is a bilinear function from $K\times K$ to $L$. These data describe a group-like braided category whose fusion rules are encoded by the abelian group $K$ while the automorphism group of the identity object is $L$. The associator is trivial, while the braiding is given by the map $\{\cdot,\cdot\}$. 

Another interesting special case is when the middle homology of the 2-crossed module vanishes, i.e. $\pi_2(\bH)$ is trivial. Then the  corresponding homotopy 3-type is completely determined by $\pi_1(\bH)=\coker\partial,$ $\pi_3(\bH)=\ker\delta$, and a Postnikov class ${\mathfrak t}\in H^4(\pi_1(\bH),\pi_3(\bH))$. An explicit construction of a 4-cocycle representing this class from the data of a 2-crossed module is given in \cite{Conduche1}.

\subsection{A crossed square for a 2d lattice system}

Consider now a 2d quantum system. Given a choice of a half-plane $\HP$, we define a crossed square as follows:
\begin{align}
L=\SU_l, M=QCA_L, N=QCA_R, P=QCA^c_{\HP}.
\end{align}
The homomorphisms forming the square are the obvious ones ($f(u)=Ad_u$, $g(u)=Ad_u$, etc.) The map $\eta$ is given by the commutator pairing defined in Section 3. It is straightforward to check that all the properties are satisfied. The homotopy groups of the 3-group $\bS_{\HP}$ corresponding to this crossed square are
$$
\pi_1(\bS_{\HP})=QCA^c_{\HP}/QCA^c_{\partial\HP},\quad \pi_2(\bS_{\HP})=1,\quad \pi_3(\bS_{\HP})=U(1).
$$
Hence its isomorphism class is completely determined by $\pi_1(\bS_\HP),\pi_3(\bS_\HP),$ and a class ${\mathfrak t}\in H^4(\pi_1,\pi_3)$. The 3-group $\bS_\HP$ describes kinematic symmetries localized to the half-plane $\HP$. 

Now suppose a group $G$ acts on the 2d system by circuits. This means that we are given a homomorphism $\rho_0:G\ra QCA^c_{\RR^2}$. Truncating it to the half-plane, we get a well-defined homomorphism $\rho:G\ra QCA^c_{\HP}/QCA_{\partial\HP}$. As discussed above, to every such homomorphism one can attach a class in $H^2(G,\QQ_+)$ which measures an obstruction for lifting this homomorphism to a homomorphism $\rho:G\ra QCA^c_{\HP}/QCA^c_{\partial\HP}=\pi_1(\bS_{\HP})$. Suppose this obstruction vanishes, and we made a choice of a homomorphism $\rho$. Then there is a well-defined class in $H^4(G,U(1))$, the pull-back of the Postnikov class in $H^4(\pi_1(\bH_\HP),U(1))$. The formula for the 4-cocycle $\tau$ in Section 4 was obtained by pulling back the 4-cocycle written down in \cite{Conduche1}. 

As in the 1d case, we can interpret this as follows. A group $G$ can be viewed as a 3-group whose only nontrivial homotopy group is $\pi_1$. Given a group homomorphism from $G$ to $\pi_1(\bS_{\HP})$, we can ask whether it can be lifted to a morphism of 3-groups $G\ra {\mathfrak S}_{\HP}$. The vanishing of the 2d anomaly index is a necessary and sufficient condition for the existence of such a lift. Accepting the interpretation proposed in the introduction, the 2d anomaly index measures the obstruction for the localizability of the action of $G$ to the half-plane $\HP$.

\subsection{An example}

In this section we present an example of a group action in two dimensions with a nontrivial anomaly.\footnote{This group action is considered in \cite{PRX} where it is argued that the corresponding domain walls obey a nontrivial "stringy statistics".} We take $G=\mathbb{Z}_2 \times \mathbb{Z}_2$ so that $H^4(G,U(1))=\ZZ_2\times\ZZ_2$. Each element $g \in G$ is denoted by a 2-tuple $g=(g_1,g_2)$, where $g_1,g_2 \in \{0,1\}$. To describe the nontrivial cohomology classes more explicitly, let $a,b : \mathbb{Z}_2 \times \mathbb{Z}_2 \to 
\ZZ_2$ be the natural projections defined by
\begin{align}
    a(g) &= (-1)^{g_1}, \\
    b(g) &= (-1)^{g_2}.
\end{align}
These are group homomorphisms and hence represent elements of 
$H^1(\mathbb{Z}_2 \times \mathbb{Z}_2,\ZZ_2)$. Therefore we get classes $a^3b$ and $b^3a$ in $H^4(\mathbb{Z}_2 \times \mathbb{Z}_2,\ZZ_2)$. One can show that the images of these classes in $H^4(\mathbb{Z}_2 \times \mathbb{Z}_2,U(1))$ are generators.

\begin{figure}[h]
    \centering
    \includegraphics[width=0.4\textwidth]{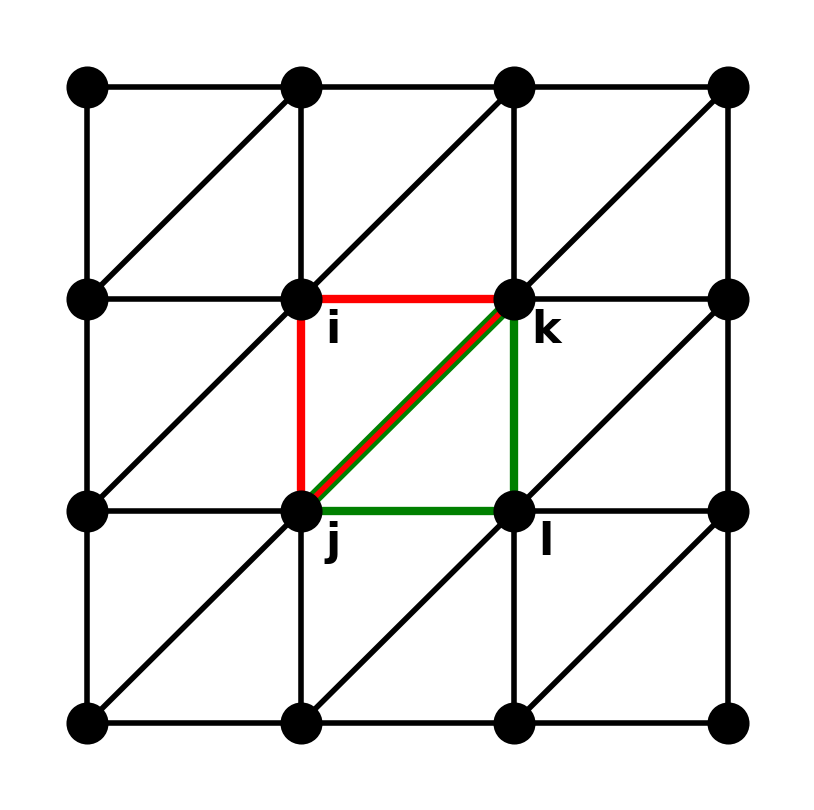}
    \caption{$CCZ_{ijk}$ is supported on the red triangle and $CCZ_{jkl}$ is supported on the green triangle}
    \label{fig:lattice}
\end{figure}

We work on a 2-dimensional square lattice with one qubit at each vertex.  Consider the following two unitaries:
\begin{align}
U_1&=\prod_{\langle ijkl \rangle}CCZ_{ijk}CCZ_{jkl} \\
U_2&=\prod_{i}X_i
\end{align}
The gate $CCZ_{ijk}$ is the controlled-controlled-$Z$ acting on qubits $i,j,k$, which multiplies the basis state $\lvert a_i a_j a_k \rangle$ by $(-1)^{a_i a_j a_k}$. The product in $U_1$ is taken over all squares of the lattice, with $(i,j,k,l)$ denoting the four vertices. More precisely, $ijk$ and $jkl$ correspond to the top-left and bottom-right triangles of each square, respectively. Equivalently, one can view the lattice as triangulated by adding the diagonal $jk$ to each square. In this picture, $U_1$ is simply the product of $CCZ$ gates over all triangles of the lattice, see Figure \ref{fig:lattice}. 

Both of these unitaries square to the identity, since each individual $CCZ$ and $X$ gate squares to the identity. To see that they commute, recall the following identity:
    \begin{align}
& X_i  C C Z_{i j k}=C Z_{j k} \ (C C Z_{i j k} X_i) \\
& X_j  C C Z_{i j k}=C Z_{k i} \ (C C Z_{i j k} X_j)\\
& X_k  C C Z_{i j k}=C Z_{i j} \ (C C Z_{i j k} X_k)
\end{align}

Swapping a $CCZ$ gate with an overlapping $X$ gate produces an extra $CZ$ gate on the non-overlapping edge. The gate $CZ_{ij}$ is the controlled-$Z$ acting on qubits $i,j$, which multiplies the basis state $\lvert a_i a_j \rangle$ by $(-1)^{a_i a_j}$. If we swap all $X$ gates with all $CCZ$ gates, each extra $CZ$ gate appears exactly twice on every edge and therefore cancels out. Hence $U_1$ commutes with $U_2$, and together they generate a $\mathbb{Z}_2 \times \mathbb{Z}_2$ symmetry. For each $g \in G$, we define
\begin{align}
\rho(g) = U_1^{g_1} U_2^{g_2}.
\end{align}

Now let $H$ denote the upper half of the lattice, and define the restriction to $H$ by

\begin{align}
\tilde{\rho}(g)=
\biggl(\prod_{\langle ijkl \rangle \in H} CCZ_{ijk}\, CCZ_{jkl}\biggr)^{g_1}
\cdot
\biggl(\prod_{i \in H} X_i\biggr)^{g_2}.
\end{align}

The first step is to compute
\begin{align}
\mu(g,h) = \tilde{\rho}(g)\,\tilde{\rho}(h)\,\tilde{\rho}(gh)^{-1}.
\end{align}
This amounts to identifying the extra terms generated when swapping 
$\bigl(\prod_{i \in H} X_i\bigr)^{g_2}$ and 
$\bigl(\prod_{\langle ijkl \rangle \subseteq H} CCZ_{ijk} CCZ_{jkl}\bigr)^{h_1}$ 
using the identity above. Away from the boundary, each extra $CZ$ term appears exactly twice on an edge and cancels. Thus, only the $CZ$ terms on the boundary edges remain. We obtain
\begin{align}
   \mu(g,h) = \left(\prod_{\langle ij \rangle \in \partial H} CZ_{ij}\right)^{g_2 h_1}. 
\end{align}

For the second step, let $R$ denote the right side of $\partial H$ starting from the origin. We define the restriction of $\mu(g,h)$ to $R$ in the natural way:
\begin{align}
    \beta(g,h) = \left(\prod_{\langle ij \rangle \in R} CZ_{ij}\right)^{g_2 h_1}.
\end{align}
Now we compute
\begin{align}
\operatorname{Ad}_{u(g,h,k)} = \beta(g,h)\,\beta(gh,k)\,\beta(g,hk)^{-1}\,{}^{\tilde{\rho}(g)}\!\beta(h,k)^{-1}.
\end{align}
All of the pure $\beta$ terms consist only of $CZ$ gates, which commute and therefore cancel. Thus, the only nontrivial contribution comes from the term ${}^{\tilde{\rho}(g)}\!\beta(h,k)^{-1}$.

\begin{align}
{}^{\tilde{\rho}(g)} \beta(h, k)^{-1}
= \biggl(\prod_{i} X_i \biggr)^{g_2}
  \biggl(\prod_{\langle ij \rangle \in R} CZ_{ij} \biggr)^{h_2 k_1}
  \biggl(\prod_{i} X_i \biggr)^{g_2}
\end{align}

Here we use the identities
\begin{align}
X_i \, CZ_{ij} &= Z_j \, (CZ_{ij} X_i), \\
X_j \, CZ_{ij} &= Z_i \, (CZ_{ij} X_j) .
\end{align}
Swapping a $CZ$ gate with an overlapping $X$ gate produces an extra $Z$ gate on the non-overlapping vertex. For all interior vertices in $R$, exactly two such $Z$ gates appear and cancel. Thus, the only nontrivial contribution occurs at the origin:
\begin{align}
    u(g,h,k) = Z_{\mathrm{ori}}^{\,g_2 h_2 k_1}.
\end{align}


Finally, we compute $\tau(g,h,k,l)$ using eq. (\ref{eq:tau}). 
Note that since QCAs are represented as unitaries, all actions are by conjugation. Moreover, because every component is an involution, we need not distinguish between a QCA and its inverse.

Here are some of the key steps in the computation:
\begin{itemize}
    \item $\beta(a,b)\bigl(u(c,d,e)\bigr) = u(c,d,e)$.  
    Since $\beta$ consists only of $CZ$ terms, which commute with $u(c,d,e)$, this action is trivial. 

    \item ${}^{\tilde{\rho}(c)}\!\beta(a,b) = \beta(a,b)$.  
    Swapping $CZ$ and $X$ terms generates extra $Z$ gates, but as all vertices here are interior, each $Z$ gate appears twice and cancels. 

    \item $\eta(\alpha(a,b),\beta(c,d)) = 1$.  
    Recall that $\operatorname{Ad}_{\eta(\alpha,\beta)} = [\alpha,\beta]$. In this case, both $\alpha$ and $\beta$ contain only $CZ$ terms, which commute, so the commutator is trivial. 

    \item  Recall the basic commutation relation $X_i Z_i = - Z_i X_i$. Thus, swapping an $X$ and a $Z$ gate produces an extra $-1$ phase.
\end{itemize}

Now we evaluate each term individually:
\begin{enumerate}
    \item $u(g,h,k)=Z_{ori}^{g_2h_2k_1}$
    \item ${}^{\tilde{\rho}(g)} \beta(h, k)(u(g, h k, l))=  \beta(h, k)(u(g,hk,l))=u(g,hk,l)=Z_{ori}^{g_2(h_2+k_2)l_1}$
    \item$ \tilde{\rho}(g)(u(h,k,l))=
     \left (\prod_{i\in H}X_i\right)^{g_2} Z_{ori}^{h_2k_2l_1}  \left (\prod_{i\in H}X_i\right)^{g_2}=(-1)^{g_2h_2k_2l_1}Z_{ori}^{h_2k_2l_1}$
     \item ${}^{\tilde{\rho}(g) \tilde{\rho}(h) }\beta(k, l)\left(u(g, h, k l)^{-1}\right)=\beta(k, l)\left(u(g, h, k l)^{-1}\right)=u(g,h,kl)=Z_{ori}^{g_2h_2(k_1+l_1)}$
     \item $\eta(\alpha(g, h), {}^{\beta(g, h) \tilde{\rho}(g h)} \beta(k, l)) =\eta(\alpha(g,h),\beta(k,l))=1$
     \item $\beta(g, h)\left(u(g h, k, l)^{-1}\right)=u(gh,k,l)=Z_{ori}^{(g_2+h_2)k_2l_1}$
    \end{enumerate}
Combining the terms, we obtain
\begin{align}
\tau(g,h,k,l)=(-1)^{g_2h_2k_2l_1}.
\end{align}
On the other hand,
\begin{align}
    (b \smile b \smile b \smile a)(g,h,k,l) = (-1)^{g_2 h_2 k_2 l_1},
\end{align}
represents a nontrivial class in $H^4(\mathbb{Z}_2 \times \mathbb{Z}_2; U(1))$. Hence, the symmetry action indeed gives rise to a nontrivial anomaly. In \cite{PRX}, this is interpreted as a nontrivial statistics for the corresponding domain walls. Ref. \cite{PRX} proposes a  different way to detect string statistics. It would be interesting to understand the connection between their methods and ours. 

\section{SPT states and higher symmetry}

In this section we show that the SPT invariants of short-range entangled states can also be understood in terms of higher groups and their morphisms. 

We will say that a state $\omega$ of a quantum lattice system is short-range entangled if it has the form $\omega_0\circ\alpha$, where $\omega_0$ is a pure product state and $\alpha$ is a QCA generated by a circuit.

Suppose we are given an action of $G$ on a 1d system by circuits and a $G$-invariant short-range entangled state $\omega$. Then the Nayak-Else index of the $G$-action is trivial \cite{NayakElse}. We recall the argument from \cite{NayakElse}. The values of $\omega\circ\trho(g)$ and $\omega$ on all local observables supported sufficiently far from the origin are equal, and this implies that for any $g\in G$ there exists  $U(g)\in\SU_l$ such that $\omega\circ \Ad_{U(g)}\circ\trho(g)=\omega$. Thus we can choose the QCAs $\trho(g)$ so that they preserve the state $\omega$, and then the combination
\begin{align}
    \mu(g,h)=\trho(g)\trho(h)\trho(gh)^{-1}\in QCA^c_0
\end{align}
preserves $\omega$ for all $g,h\in G$. Therefore the equation 
\begin{equation}
    \Ad_{V(g,h)}=\mu(g,h)
\end{equation}
is solved by $V(g,h)\in \SU_l$ which preserves $\omega$ up to a phase and thus satisfies $\omega(V(g,h))\in U(1)$ for all $g,h\in G$. It is easy to see that the 2-cochain $c(g,h)=\omega(V(g,h))$ is a trivialization of the Nayak-Else class, i.e. $\rho^*{\mathfrak l}=\delta c.$ Here $\rho:G\ra QCA_R/QCA_0$ describes the action of $G$ truncated to a half-line. 

We can interpret this in terms of higher groups symmetry as follows. For any state $\omega$ of a 1d lattice system we may consider a 2-group $\bH_R^\omega$ of symmetries of $\omega$ localized to $R$. The corresponding crossed module has the form $\SU_l^\omega\ra QCA_R^\omega$, where $QCA_R^\omega$ is the group of QCAs localized on R and preserving $\omega$ and $\SU_l^\omega\subset\SU_l$ consists of elements of $\SU_l$ which are mapped by $\Ad$ to $QCA_R^\omega$. This 2-group is a subgroup of $\bH_R$, and thus $\pi_1(\bH^\omega_R)$ is a subgroup of $\pi_1(\bH_R)$. If the state $\omega$ is a $G$-invariant state, then the image of the homomorphism $\rho:G\ra \pi_1(\bH_R)$ lands in this subgroup, and one may ask if it can be lifted to a 2-group morphism $G\ra \bH^\omega_R$. The above arguments show that this can always be done if $\omega$ is a $G$-invariant SRE state. Composing this 2-group morphism with the embedding $\bH_R^\omega\ra\bH_R$, we get a morphism $G\ra\bH_R$, which implies that the Nayak-Else class is trivial. Put in more physical terms, an action of $G$ by circuits which preserves an SRE state is always localizable to a half-line. 

If we are given two $G$-invariant SRE states $\omega$ and $\omega'$, then we get two trivializations $c$ and $c'$ whose difference satisfies
$\delta(c {c'}^{-1})=0$, i.e. it is a 2-cocycle on $G$ with values in $U(1)$. The cohomology class of this 2-cocycle is independent of the choice of truncations $\trho(g)$ and $\trho'(g)$ preserving $\omega$ and $\omega'$. Indeed, any two such truncations preserving $\omega$ (resp. $\omega'$) differ by a QCA of the form $\Ad_{W(g)}$, and this changes $c$ (resp. $c'$) by the coboundary of the $U(1)$-valued 1-cochain $\omega(W(g))$. 

The class of the 2-cocycle $c {c'}^{-1}$ is a relative invariant of the $G$-invariant SRE states $\omega$ and $\omega'$. Typically, this invariant is discussed in the situation when $G$ acts on-site and is non-projective (i.e. its restriction to observables on any site arises from a homomorphism from $G$ to the unitary group of the on-site Hilbert space). Then there is an almost canonical choice for $\omega'$ (namely, a $G$-invariant product state), and the relative invariant of the pair $(\omega,\omega')$ may be regarded as an absolute invariant of $\omega$. If it is nontrivial, $\omega$ is a nontrivial Symmetry Protected Topological (SPT) state.

From the higher-group viewpoint, given a pair of $G$-invariant SRE states $\omega,\omega'$, we get a pair of 2-group morphisms $G\ra\bH_R$. Recalling the discussion in Section 5, we see that if the relative SPT class is nontrivial, then these two morphisms are not isomorphic. In homotopy-theoretic terms, the relative SPT index measures the difference between homotopy classes of maps from $BG$ to the 2-group $\bR$ corresponding to the two states $\omega$ and $\omega'$. 

In physical terms, when $G$ preserves an SRE state, its action can be promoted to a local action (which also preserves the state), but different $G$-invariant SRE states may give rise to inequivalent local actions. The SPT invariant is a measure of this inequivalence. This is in harmony with the interpretation of SPT invariants in QFT, where they are identified as "topological" (i.e. quantized) ambiguities in the contact terms in the current correlators, or equivalently as topological terms for the background gauge fields. 

So far we discussed the case of 1d SRE states. The situation in 2d is similar: if a $G$-invariant SRE state exists, it trivializes the 2d anomaly. To see this, let $\trho:G\ra QCA^c_{\HP}$ be a truncation of a homomorphism $\rho_0:G\ra QCA^c_{\RR^2}$ which preserves a 2d SRE state $\omega$. Then $\omega^{g}=\omega\circ\trho(g)$ differs from $\omega$ only in some thickening of $\partial\HP$. It is not hard to show that this implies $\omega^g\circ\kappa(g)=\omega$ for some $\kappa(g)\in QCA^c_{\partial\HP}$. Indeed, since $\omega$ can be disentangled by a circuit, without loss of generality we may assume that $\omega$ is a factorized pure state, and thus $\omega^g$ is a 1d pure state on some thickening of $\partial\HP$. Moreover, it is clear that $\omega^g$ has a finite range of correlations, and therefore is a 1d SRE state. $\kappa(g)$ is a disentangler of this state. Therefore, by redefining $\trho(g)$, we can make it $\omega$-preserving. The same is true about $\beta(g,h)\in QCA_R$ and $u(g,h,k)\in\SU_l$. Then  $\omega(u(g,h,k))$ is a 3-cochain with values in $U(1)$, and one can show that it is a trivialization of the 4-cocycle $\tau$. Conversely, if the 2d anomaly index of a $G$-action is nontrivial, no SRE state can be invariant under this action. 

It was proved in \cite{KapSop} that a nonzero Nayak-Else index also obstructs the existence of $G$-invariant 1d gapped ground states. By analogy, one might conjecture that a nonzero 2d anomaly index obstructs the existence of $G$-invariant invertible 2d states. We remind that a pure state is called invertible if its tensor product with some other pure state can be created from an unentangled state by a sufficiently nice automorphism (for example, a Locally Generated Automorhism \cite{LocalNoether}).

\section{Discussion}

It is well appreciated that 't Hooft anomalies are obstructions for promoting a global symmetry $G$ to a local symmetry. In this paper we made this mathematically precise in the case of lattice systems in one and two spatial dimensions by interpreting locality in homotopy-theoretic terms. Namely, locally acting symmetries are described by a higher group, and an 't Hooft anomaly is an obstruction for lifting a homomorphism of ordinary groups to a weak morphism of higher groups. Such obstructions can be nontrivial only when the higher group is not merely a product of ordinary symmetry and higher-form symmetry.

We also offered a new mathematical interpetation of invariants of SPT states. Namely, each $G$-invariant SRE state gives a preferred way to promote a global $G$-action to a local one, but different states may give rise to inequivalent ways of doing so. Homotopy theory is helpful for defining a proper notion of equivalence of local symmetry actions. 

Finally, we observed that homotopy theory enters the study of lattice systems through its algebraic incarnations such as crossed modules and crossed squares. This appears to be an important lesson not just for lattice models but for QFT too. It remains to be seen which algebraic models of homotopy types are the most natural ones from the point of view of physical applications.

\appendix

\section{Some computations}

Let us show that $\Ad_{\tau(g,h,k,l)}=1$. We compute

\begin{align}
\Ad_{u(g,h,k)}=\beta(g,h)\beta(gh,k)\beta(g,hk)^{-1}\trho(g)\beta(h,k)^{-1}\trho(g)^{-1},
\end{align}
\begin{multline}
\trho(g)\beta(h,k)\trho(g)^{-1}(\Ad_{u(g,hk,l)})=\trho(g)\beta(h,k)\trho(g)^{-1}\beta(g,hk)\beta(ghk,l)\\\beta(g,hkl)^{-1}
\trho(g)\beta(hk,l)^{-1}\trho(g)^{-1}
\trho(g)\beta(h,k)^{-1}\trho(g)^{-1}=\\
=\trho(g)\beta(h,k)\trho(g)^{-1}\beta(g,hk)\beta(ghk,l)\beta(g,hkl)^{-1}
\\
\trho(g)\beta(hk,l)^{-1}\beta(h,k)^{-1}\trho(g)^{-1},
\end{multline}
\begin{multline}
\Ad_{\trho(g)(u(h,k,l))}=\trho(g) \beta(h,k)\beta(hk,l)\beta(h,kl)^{-1}\\
\trho(h)\beta(k,l)^{-1}\trho(h)^{-1}\trho(g)^{-1},
\end{multline}
\begin{multline}
\Ad_{\trho(g)\trho(h)\beta(k,l)\trho(h)^{-1}\trho(g)^{-1}(u(g,h,kl)^{-1})}=\\
\trho(g)\trho(h)\beta(k,l)\trho(h)^{-1}\trho(g)^{-1}\trho(g)\beta(h,kl)\trho(g)^{-1}\beta(g,hkl)\beta(gh,kl)^{-1}\beta(g,h)^{-1}\\
\trho(g)\trho(h)\beta(k,l)^{-1}\trho(h)^{-1}\trho(g)^{-1}=\\
=\trho(g)\trho(h)\beta(k,l)\trho(h)^{-1}\beta(h,kl)\trho(g)^{-1}
\beta(g,hkl)\beta(gh,kl)^{-1}\beta(g,h)^{-1}\\
\trho(g)\trho(h)\beta(k,l)^{-1}\trho(h)^{-1}\trho(g)^{-1},
\end{multline}
\begin{multline}
\left[\alpha(g,h),\beta(g,h)\trho(gh)\beta(k,l)\trho(gh)^{-1}\beta(g,h)^{-1}\right]=\\
\left[\trho(g)\trho(h)\trho(gh)^{-1}\beta(g,h)^{-1},\beta(g,h)\trho(gh)\beta(k,l)\trho(gh)^{-1}
\beta(g,h)^{-1}\right]=\\
\trho(g)\trho(h)\beta(k,l)\trho(gh)^{-1}\beta(g,h)^{-1}\beta(g,h)\trho(gh)\trho(h)^{-1}\trho(g)^{-1}\beta(g,h)\trho(gh)\\
\beta(k,l)^{-1}\trho(gh)^{-1}\beta(g,h)^{-1}=\\
=\trho(g)\trho(h)\beta(k,l)\trho(h)^{-1}\trho(g)^{-1}\beta(g,h)\trho(gh)\beta(k,l)^{-1}\trho(gh)^{-1}\beta(g,h)^{-1},
\end{multline}
\begin{multline}
\Ad_{\beta(g,h)(u(gh,k,l)^{-1})}=\beta(g,h)\trho(gh)\beta(k,l)\trho(gh)^{-1}\beta(gh,kl)\\
\beta(ghk,l)^{-1}\beta(gh,k)^{-1}\beta(g,h)^{-1}.
\end{multline}

Multiplying the first three expressions we get
\begin{multline} \beta(g,h)\beta(gh,k)\beta(ghk,l)\beta(g,hkl)^{-1}\trho(g)\beta(h,kl)^{-1}\\
\trho(h)\beta(k.l)^{-1}\trho(h)^{-1}\trho(g)^{-1}.
\end{multline}
Multiplying the last three expressions we get
\begin{multline}  \trho(g)\trho(h)\beta(k,l)\trho(h)^{-1}\beta(h,kl)\trho(g)^{-1}
\beta(g,hkl)\\
\beta(ghk,l)^{-1}\beta(gh,k)^{-1}\beta(g,h)^{-1}.
\end{multline}
Multiplying the last two expressions, we get $1$.

Now suppose we replace the function $\beta:G\times G$ with a function $\beta'=\Ad_{v}\circ\beta$, where $v:G\times G\ra \SU_l$ is arbitrary. Let 
\begin{multline}
u'(g,h,k)=v(g,h)\beta(g,h)(v(gh,k)) u(g,h,k) {}^{\trho(g)}\!\beta(h,k)(v(g,hk)^{-1})\\
\trho(g)(v(h,k)^{-1}).
\end{multline}
It is easy to verify that $u'(g,h,k)$ solves
\begin{align}
\Ad_{u'(g,h,k)}=\beta'(g,h)\beta'(gh,k)\beta'(g,hk)^{-1}\trho(g)(\beta'(h,k)^{-1}).
\end{align}
A computation which is too lengthy to include here shows that replacing $u$ with $u'$ does not affect the 4-cochain $\tau$. 

Finally, suppose we replace $\trho:G\ra QCA^c_{\HP}$ with
$\trho'=\gamma\circ\trho$, where $\gamma:G\ra QCA^c_{\partial\HP}$ is an arbitrary function. Decomposing $\gamma=\gamma_L\gamma_R$ where $\gamma_R:G\ra QCA_R$ and $\gamma_L:G\ra QCA_L$, we see that $\beta$ is replaced with
\begin{align}
\beta'(g,h)=\gamma_R(g)\,{}^{\trho(g)}\!\gamma_R(h)\beta(g,h)\gamma_R(gh)^{-1}.
\end{align}
This changes $u$ to
\begin{multline}
u'(g,h,k)=\gamma_R(g)\,{}^{\trho(g)}\!\gamma_R(h)\left(\eta\left(\alpha(g,h),{}^{\beta(g,h)\trho(gh)}\!\gamma_R(k)\right)\right)\gamma_R(g)\,{}^{\trho(g)}\!\gamma_R(h)\\
{}^{\trho(g)\trho(h)}\!\gamma_R(k)\left(u(g,h,k)\right)\eta\left(\gamma_L(g),{}^{\gamma_R(g)\trho(g)}\!\beta'(h,k)\right).
\end{multline}
A long computation shows that these changes do not affect the 4-cochain $\tau$.

\bibliographystyle{unsrt} 
\bibliography{Bibliography.bib} 

 \end{document}